# Twist-Induced Hyperbolic Shear Metasurfaces

Simon Yves[1$], Emanuele Galiffi[1$], Xiang Ni[1], Enrico Maria Renzi[1], Andrea Alù[1,2*]

[1]Photonics Initiative, Advanced Science Research Center, City University of New York, New York, NY 10031, USA

[2]Physics Program, Graduate Center, City University of New York, New York, NY 10026, USA

[*]To whom correspondence should be addressed (email: aalu@gc.cuny.edu)

[$] Contributed equally to this work

Following the discovery of moiré-driven superconductivity[1] and density waves[2] in twisted graphene multilayers, twistronics[3] has spurred a surge of interest in tailored broken symmetries through angular rotations enabling new properties, from electronics[4-7] to photonics[8-17] and phononics[18-26]. Analogously, in monoclinic polar crystals a nontrivial angle between non-degenerate dipolar phonon resonances can naturally emerge due to asymmetries in their crystal lattice, and its variations are associated with intriguing polaritonic phenomena, including axial dispersion, i.e., the rotation of the optical axis with frequency, and microscopic shear effects that result in asymmetric distribution of material loss[27,28]. So far these phenomena have been restricted to specific mid-infrared frequencies, difficult to access with conventional laser sources, and fundamentally limited by the degree of asymmetry and by the strength of light-matter interactions available in natural crystals. Here, we leverage twistronic concepts to demonstrate giant axial dispersion and loss asymmetry of hyperbolic waves in elastic metasurfaces, achieved by tailoring the angle between coupled metasurface pairs featuring tailored anisotropy. We show extreme control over elastic wave dispersion via the twist angle, and leverage the resulting phenomena to demonstrate in-plane reflection-free negative refraction, as well as the application

**of axial dispersion to achieve diffraction-free non-destructive testing, whereby the angular direction of a hyperbolic probe wave is encoded into its frequency. Our work welds the concepts of twistronics, non-Hermiticity and extreme anisotropy, demonstrating the powerful opportunities enabled by metasurfaces for tunable, highly directional surface acoustic wave propagation, of great interest for a wide range of applications spanning from seismic mitigation to on-chip phononics and wireless communication systems, hence paving the way towards their translation into emerging photonic and polaritonic metasurface technologies.**

Breaking symmetries is paramount to achieving exquisite control over light and sound propagation. A landmark example is found in hyperbolic waves, characterized by strongly directional, ray-like propagation[29-32] driven by extreme asymmetry in the material response. Such atypical wave dynamics can naturally emerge in polar dielectrics that support hybrid light-matter quasiparticles stemming from the resonant coupling between photons and strong in-plane anisotropic lattice vibrations, namely hyperbolic phonon polaritons[33-39]. These peculiar surface waves exhibit extreme field confinement, ultralow loss, and highly canalized propagation at mid-infrared frequencies, offering exciting opportunities for superior light manipulation, large light-matter coupling, sensing and imaging at the nanoscale.

In this context, a further degree of asymmetry has been recently unveiled in monoclinic polar crystals, where a new form of surface polaritons has been demonstrated to emerge from the interaction of light with two non-degenerate dipolar phonon resonances whose orientation forms a non-trivial angle. The resulting *hyperbolic shear polaritons* are characterized by microscopic shear phenomena that lead to an in-plane rotation of their optical axis with frequency, accompanied by asymmetric damping of the supported hyperbolic waves[27,28]. Compounding the intrinsic directionality of hyperbolic waves with this loss-asymmetry leads to a new degree of wave control in nano-optics. While exciting from a fundamental level, natural material platforms supporting hyperbolic shear polaritons suffer from several drawbacks, such as the inherent limitation to mid-infrared frequencies for the relevant phonon resonances, difficult to access with commercial lasers, lack of tunability, and restrictions on the degree of asymmetry naturally available in crystal lattices. These constraints prevent a deeper manipulation and control of shear-hyperbolic responses, restraining their range of potential applications.

To overcome these limitations and harness hyperbolic shear waves on-demand, here we extend this concept to hyperbolic metasurfaces, structured thin sheets patterned with subwavelength resonator arrays that realize strong in-plane anisotropy, enabling precise control over the symmetries that govern wave propagation, and dramatically enhancing wave-matter interactions[40-43]. As a new paradigm to design hyperbolic shear metasurfaces, given the important role of rotations and asymmetries, it appears natural to embrace twistronics[3] from condensed-matter systems, based on which the combined rotation and stacking of rotated layered materials can induce exotic responses, such as superconductivity[1] and density waves[2]. Without loss of generality, here we work with elastodynamic waves, overlaying two thin elastic 3D-printed metasurfaces, each loaded with a different subwavelength array of anisotropic pillar resonators[44]. By leveraging their reconfigurable twist angle, we achieve precise control over shear wave phenomena, demonstrating giant axial dispersion and mirror-asymmetric lifetime of flexural hyperbolic shear waves. Finally, we leverage the unique properties of hyperbolic shear waves to demonstrate reflectionless negative refraction at an interface, and propose a technological avenue for this concept to realize axial dispersion-based surface scanning for diffraction-free non-destructive testing, as a forerunner to future sonar and radar technology.

Our results establish an intuitive, general paradigm for the realization and control of twist-governed monoclinic hyperbolic metasurfaces extendable across multiple wave domains. Beyond the experimental demonstration of hyperbolic shear phenomena, our elastodynamic platform paves the way for advanced surface-wave technologies ranging from on-chip phononics to seismic mitigation, as well as metasurface-based devices for radio-frequency telecommunications and nanophotonic technologies.

**Monoclinic hyperbolic metasurfaces**

Consider an elastic metasurface formed by an array of subwavelength unit cells, each featuring two dipolar resonances $R_1$ and $R_2$. Their respective Lorentzian response functions $\tau_1(\omega)$ and $\tau_2(\omega)$ account for their hybridization with the flexural waves of a thin plate (equivalently to the dielectric permittivity of a polaritonic material), for an excitation oriented along their oscillator directions. If the two resonators are orthogonal, the resulting linear response tensor $\hat{\tau}$ is uniaxial, with different diagonal components $\tau_{xx} = \tau_1$ and $\tau_{yy} = \tau_2$, and null off-diagonal terms $\tau_{xy} = \tau_{yx} = 0$, defined in a Cartesian reference frame whose axes $x$ and $y$ are aligned with the resonator symmetry axes (Fig. 1a, left). In the absence of absorption, the elements of $\hat{\tau}$ are Hermitian and, if the two resonances are non-degenerate, a band between the two resonance frequencies emerges, within which $\tau_{xx}$ and $\tau_{yy}$ have opposite signs. In this frequency range, the metasurface supports hyperbolic elastic waves, in analogy to the hyperbolic polaritons demonstrated in various uniaxial polar crystals in their Reststrahlen bands[37,39].

Assuming a general Kirchoff-Love model for flexural waves supported by an elastic plate in the local small-wavevector limit (homomorphic to a 2D model for the electromagnetic scalar potential, see Methods), we derive a homogeneous model that accounts for the interaction of the resonator with the flexural displacement field of the plate, as detailed in the Supplementary Information (SI), Sec. I and II. The eigenmodes of the system form hyperbolic isofrequency contours (IFCs) in momentum space, with principal axis (the symmetry axis crossing the tip of the hyperbola) aligned with the resonators (the dotted line in the right panel of Fig. 1a accounts for the three overlapping axes), but frequency-dependent asymptotes (blue to green contours in Fig. 1a), consistent with the open angle dispersion of hyperbolic phonon polaritons in uniaxial crystals[37,39].

Consider now non-orthogonal resonators, e.g., $R_1$ is aligned with the $x$-axis, while $R_2$ forms an angle $\theta$ with it, (Fig. 1b). The linear response tensor becomes (see SI Sec. I)

$$\hat{\tau} = \begin{pmatrix} \tau_1 + \tau_2 \cos^2(\theta) & -\tau_2 \sin(\theta)\cos(\theta) \\ -\tau_2 \sin(\theta)\cos(\theta) & \tau_2 \sin^2(\theta) \end{pmatrix}. \tag{1}$$

The non-zero off-diagonal terms are responsible for the interaction between the two resonators, due to their non-orthogonality. This leads to axial dispersion, namely a frequency-dependent rotation of the principal axis (dotted lines on Fig. 1b) and of the hyperbolic IFCs, by an angle (see Methods for details)

$$\beta = \frac{1}{2}\tan^{-1}\left[\frac{\Re[\tau_2]\sin(2\theta)}{\Re[\tau_1] + \Re[\tau_2]\cos(2\theta)}\right], \tag{2}$$

obtained by diagonalizing $\hat{\tau}$, consistent with the recently observed axial dispersion of hyperbolic shear polaritons in monoclinic crystals[27,28].

Once material losses are included in the oscillator responses $\tau_1(\omega)$ and $\tau_2(\omega)$, $\hat{\tau}$ becomes non-Hermitian, and the power loss rate (color of analytic IFC in Fig. 1c, see Methods and SI Sec. III for details) of the hyperbolic modes increases for larger momenta, as the modes become more localized. For orthogonal oscillators ($\theta = 90°$), the loss distribution inherits the symmetry axis of the contour (dotted line in Fig. 1c), indicating that mirror-symmetric modes are equally long-lived. Hence, a point-source excites the four hyperbolic branches with equal efficiency, yielding mirror-symmetric features in reciprocal and real space, as shown in Figs. 1d and e, respectively. By stark contrast, in a monoclinic system ($\theta \neq 90°$) the principal axis associated with the Hermitian contribution of the material response cannot be generally expected to be a symmetry axis for the dissipative processes affecting the modes. Consequently, in the frame aligned with the symmetry axis of the contour, only the real part of the tensor $\hat{\tau}' = \hat{R}(\beta)\hat{\tau}\hat{R}(\beta)^T$ is diagonal, resulting in purely imaginary off-diagonal elements $\tau'_{xy}$ in the rotated frame. This quantity measures the misalignment between the symmetry axes of the Hermitian and non-Hermitian components of the tensor, $\Re[\hat{\tau}]$ and $\Im[\hat{\tau}]$, which leads to a mirror-asymmetric distribution of the loss along the contours, exemplified in Fig. 1f (see Methods and SI Sec. III), consistent with recent observations for hyperbolic shear polaritons[27,28]. Normalizing $\tau'_{xy}$ by the total loss in the system yields the shear factor $S(\omega,\theta)$ (see Methods for details). Despite being agnostic to the IFCs of the system, the shear factor is proportional to the difference between the power loss rate calculated at any two mirror-symmetric $k$-points of a given contour, making it the relevant criterion for quantifying the strength of loss asymmetry (see SI Sec. III for details). Geometrically, this effect is associated with the misalignment between the eigenfields mirrored with respect to the principal axis (arrows in Fig. 1c,f and insets) and the eigenvectors of the non-Hermitian component of $\hat{\tau}$ (orange and purple segments in Fig. 1c,f insets denote high-loss and low-loss eigenstates of $\Im[\hat{\tau}]$ respectively).

Due to this broken symmetry, a point-source will selectively excite the two low-loss hyperbolic branches, as evident from Fig. 1g, producing a remarkably skewed, and more directional, hyperbolic wave (Fig. 1h). Notably, this angular redistribution of loss enables a dramatic extension in the lifetime of modes carrying large momenta, as opposed to the conventional hyperbolic case of Fig. 1d, despite the overall dissipation in the resonators being the same. In turn, this implies that the corresponding shear-hyperbolic waves can propagate much farther, and more directionally, than in the case of mirror-symmetric hyperbolic waves in Fig. 1e, establishing twistronics as a powerful paradigm to tailor ultralong-lived shear hyperbolic waves.

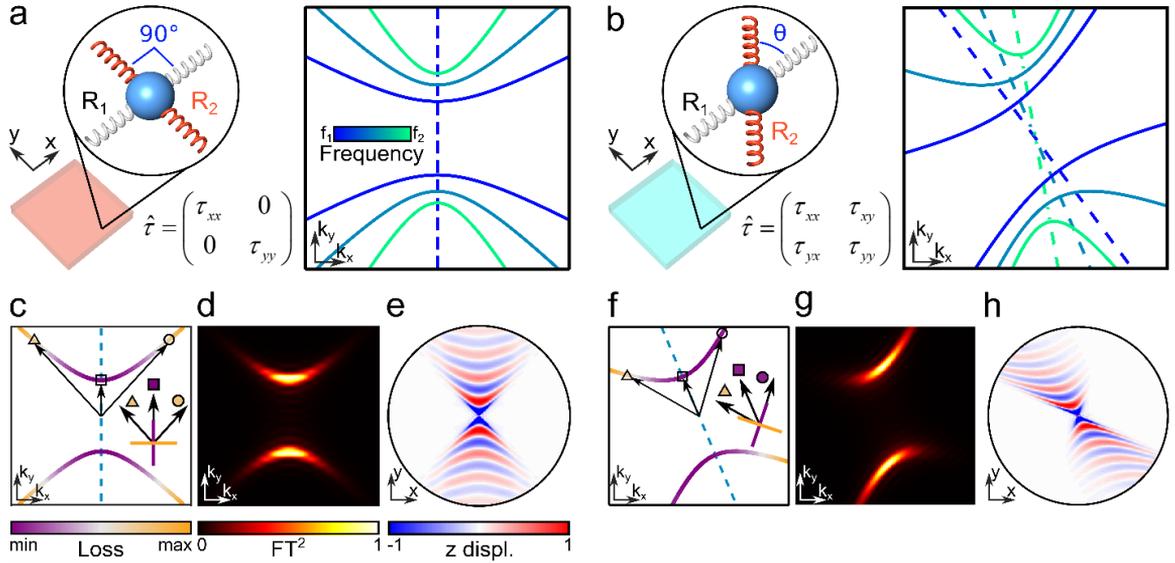

*Figure 1: Twist-induced axial dispersion and loss asymmetry. **a**, A metasurface formed by orthogonal dipolar resonators is described by a diagonal tensor $\hat{\tau}$, leading to a frequency-independent principal axis (dotted line). **b**, On the contrary, a twist angle $0° < \theta < 90°$ between the two dipolar resonators introduces off-diagonal terms in the tensor $\hat{\tau}$, resulting in axial dispersion, i.e., frequency dispersion in the principal axis orientation (dotted lines). **c,d,e**, Considering the non-Hermitian features of $\hat{\tau}$, the orthogonal scenario shown in **(a)**, leads to **(c)** conventional axisymmetric loss profiles, shown in **(d)** reciprocal and **(e)** real-space hyperbolic polariton propagation under near-field excitation. **f,g,h**, The change in relative angle between the principal axis and the underlying resonators, depicted in **(b)**, yields **(f)** asymmetric loss profiles leading to enhanced directionality in polariton propagation, shown in **(g)** reciprocal and **(h)** real space. Insets in **(c)** and **(f)** show respectively symmetric and asymmetric loss stemming from the alignment or misalignment between the eigenvectors of $\Im[\hat{\tau}]$ (orange and purple segments denote high and low-loss eigenvectors of $\Im[\hat{\tau}]$) and the field eigenvectors (black arrows) corresponding to three eigenvalues (shaped markers) located symmetrically with respect to the principal axis.*

**Twisted-bilayer elastic metasurfaces**

In order to validate our theory and demonstrate these phenomena, we fabricated a mechanical metasurface consisting of two 3D-printed thin plates, each loaded with an array of rectangular pillars (Fig. 2a, see fabrication details in Methods and Fig. S1 of SI), whose respective heights $h_1$ and $h_2$ are offset to detune their directional bending resonances. In turn, they support a hyperbolic frequency band with effective resonant material response $\Re[\tau_1(\omega)] < 0$ and $\Re[\tau_2(\omega)] > 0$ (Fig. 2b, see also SI Sec. I). The two plates are coupled using double-sided tape (inset in Fig. 2a), enabling full control over the twist angle $\theta$ between the oscillation axes of the detuned resonators in each layer.

We derive a comprehensive theory, based on the full nonlocal thin-plate Kirchoff-Love model, describing flexural wave propagation in this system through its effective linear response tensor $\hat{\tau}$ (see SI Sec. I, II for details). The resulting analytical IFCs for $\theta = 60°$ (Fig. 2c) are not open, like usual hyperbolic waves, due to the nonlocality of the Kirchoff-Love equation. Yet, they feature a hyperbolic response for small wavenumbers, and clearly show strong axial dispersion for the three frequencies denoted in Fig. 2b by blue-to-green vertical lines within the Reststrahlen band. Whilst the IFCs in the lossless scenario follow the symmetry dictated by the principal axes (dashed lines in Fig. 2c), axial dispersion alone is not sufficient to capture the simulated (Fig. 2d) phenomenology arising from point-source excitation, which is in remarkable agreement with the experimental measurements of the twisted metasurface sample (Fig. 2e). This is due to the broken mirror-symmetry in the loss distribution: as visible from the Fourier (left columns) and real (right columns) space plots, the excitation profile is very skewed, evidencing the shear features of the hyperbolic waves. This loss asymmetry closely matches our analytical prediction in Fig. 2c. Notably, this asymmetry, stemming from shear phenomena, does not hinge on the hyperbolic nature of the bands, but it arises regardless of the contour topology. Based on the previous discussion, it is due to the interplay between monoclinicity and non-Hermiticity of the effective response tensor $\hat{\tau}$ of the metasurface. Its implementation using a design as intuitive as twisted metasurfaces opens a plethora of opportunities for advanced integrable and reconfigurable devices in phononics and photonics.

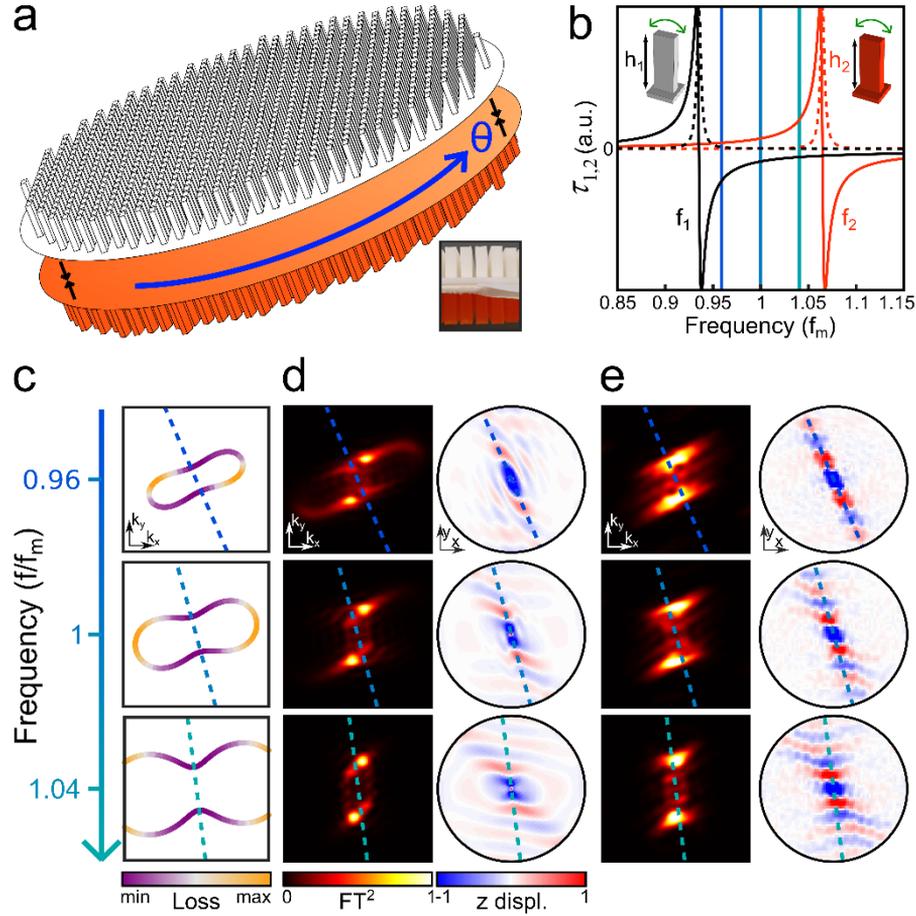

*Figure 2: Twisted elastic metasurface. **a**, Illustration of back-to-back-stacked elastic metasurfaces, twisted by an angle $\theta$. The inset shows a cross-section of the experimental sample. **b**, Frequency dependence of the effective linear response parameters $\tau_1$ and $\tau_2$ for two detuned pillar-bending resonators, of height $h_1 = 7.5$ mm and $h_2 = 7.0$ mm (inset), normalized by the center frequency $f_m = (f_1 + f_2)/2$. **c**, Analytic IFCs of homogenized elastic metasurfaces for $\theta = 60°$ and different frequencies corresponding to the blue-to-*

*green lines in (b). The contour color scale indicates the corresponding power loss rate (increasing from purple to orange). Dashed lines denote the principal axis at the three frequencies chosen in (b).* **d,** *Finite-element simulations of the flexural displacement of the pillar system under point source excitation demonstrate axial dispersion and loss asymmetry in Fourier (left) and real (right) space.* **e,** *Corresponding displacement measurements carried out with a laser vibrometer, showing excellent agreement.*

**Giant axial dispersion and shear factor**

We now quantify and maximize axial dispersion and loss asymmetry in our twisted hyperbolic shear metasurfaces. Figure 3a shows the theoretically predicted rotation angle $\beta$ as a function of frequency for different twist angles $\theta$. Notice how the axial rotation becomes increasingly abrupt as the twist-angle is reduced, enabling control over both range and rate of axial dispersion with frequency. Interestingly, the behavior of $\beta$ with respect to $\theta$ can be found in two distinct phases, depending on the dominant oscillator. At low frequencies ($|\Re[\tau_1]|>|\Re[\tau_2]|$), the principal axis fully rotates together with the twist angle between the dipolar resonators, ranging from $0°$ to $90°$. Conversely, at high frequencies ($|\Re[\tau_2]|>|\Re[\tau_1]|$), the principal axes only tilt up to a finite angle, and then revert to their original orientation along the *x*-axis (see Fig.S2 of SI for more details). The existence of this phase transition highlights the importance of the transition frequency $f \approx f_m$ (black dotted line in Fig. 3a, b), at which $|\Re[\tau_1]| \approx |\Re[\tau_2]|$. In fact, the similarity between the two detuned oscillator strengths in this transitory regime implies a stronger interaction between the two resonances, whose balanced interplay maximizes axial dispersion. This has direct consequences on the degree of loss asymmetry, as shown by the behavior of the shear factor $S$ depicted in Fig. 3b, which is indeed maximized when $f \approx f_m$ for all non-orthogonal twisted configurations, while its value increases at the expense of its bandwidth following the progressive alignment of the two resonators. This is consistent with the corresponding symmetry-breaking in the spatial distribution of losses in the medium, making the twist between metasurfaces a straightforward parameter for the precise control of both strength and bandwidth of the loss asymmetry of shear hyperbolic waves.

To showcase such versatility, we focus on the regime of maximum shear factor at frequency $f = f_m$, for which the corresponding theoretical IFCs are displayed as a function of the twist angle $\theta$ in Fig. 3c. In particular, for $\theta$ changing from $90°$ to $45°$, the contour rotation as a function of the twist angle is accompanied by a sharp increase of the loss asymmetry. As a direct consequence, strongly directional shear-hyperbolic waves exhibit a dramatic increase in propagation length along the suppressed-loss branch as the twist angle decreases, as evidenced by the theoretical field maps of Fig. 3d (top/bottom in reciprocal/real space). As the amount of dissipation in the resonators is constant throughout the twisting process, this effect emerges from the redistribution of loss induced by the finite twist-angle: while parts of the contours are overdamped, their mirrored counterparts are strongly enhanced compared to the orthogonal hyperbolic case. In reciprocal space, this effect yields a clear sharpening of the corresponding contours (Fig. 3d, top).

Twisting the metasurfaces even further ( $0° \leq \theta < 45°$ ) induces a topological transition of the contours from quasi-hyperbolic ( $\theta = 45°$ ) to flat ( $\theta = 30°$ ) and finally elliptic ( $\theta \leq 15°$ ), as the resonators progressively align. In this regime we find the maximum asymmetry in the loss distribution, corresponding to a shear factor $S \rightarrow 1$ for a passive medium, as predicted by our theory in Fig. 3b, and clearly verified by our simulations and experimental measurements in Fig. 3c-e ( $\theta = 15°$ ), where the measured signal strikingly lies entirely on one side of the principas axis. Combined with maximum loss asymmetry, this feature leads to complete control over the wave directionality through twist. These results, further corroborated by the remarkable agreement of simulations and experiments (Fig. 3e,f respectively), demonstrate that twisted hyperbolic shear metasurfaces allow for giant axial dispersion and long-range directional steering of hyperbolic waves.

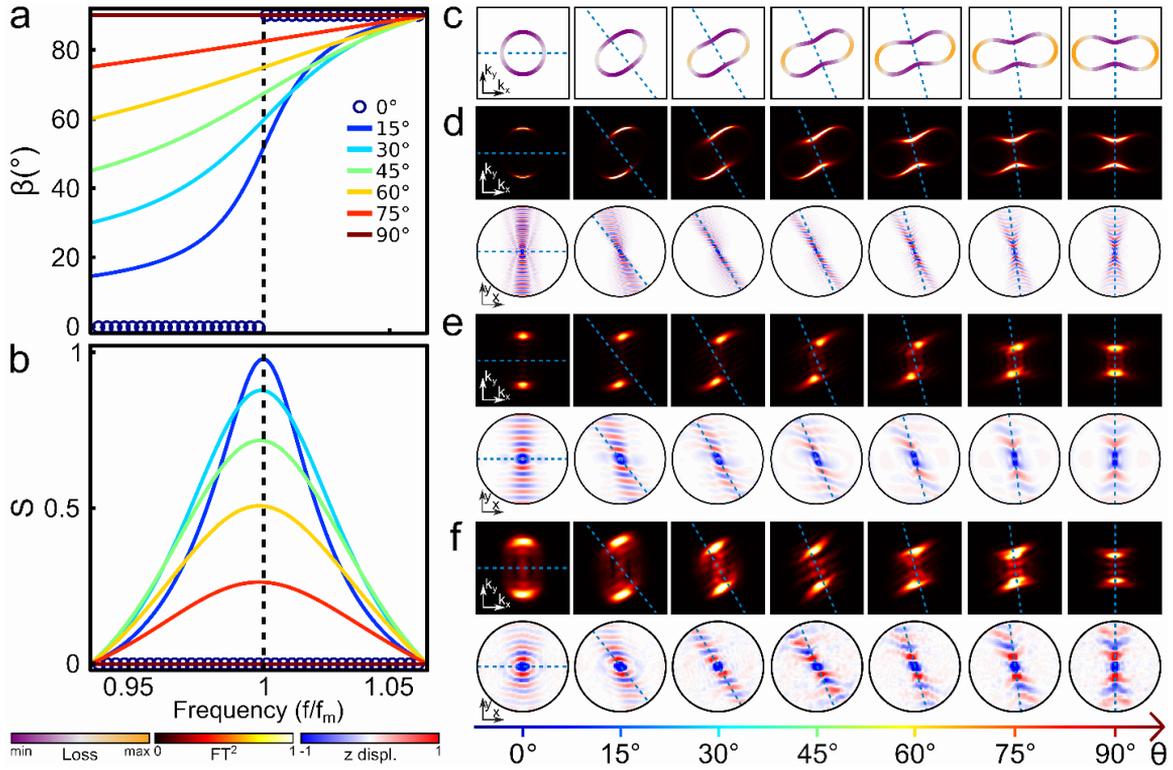

**Figure 3: Giant axial dispersion and loss asymmetry tuning with twist angle and frequency. a,** The frequency dependence of the principal axis angle $\beta$ for different twist angles $\theta$ (blue to red) reveals an axial winding phase (lower frequencies), whereby the axis fully rotates by $90°$ as $\theta$ varies from $0°$ to $90°$, and an axially stable phase (higher frequencies), whereby the principal axis rotates but then folds back. **b,** This behavior strongly impacts the bandwidth of the shear factor $S$, which peaks at the transition frequency $f_m$ (dotted black line in **a,b**), where the interaction between the two resonant modes is strongest. **c,** Twist dependence of theoretical IFCs at the frequency $f_m$ where the shear factor $S$ is maximum. **d,** Effective-medium theoretical wave propagation results both in reciprocal (top) and real (bottom) space for a large domain, corresponding to the IFCs

*displayed in (c). **e,f,** Simulation (e) and experimental counterparts of the results presented in (d) for a smaller domain. Dashed lines in (c-e) denote principal axes.*

**Reflection-free negative refraction at a hyperbolic shear interface**

The interplay between mirror-asymmetric loss distribution and hyperbolicity at a boundary offers unique opportunities to tailor hyperbolic wave propagation and scattering. As an example, it is well known that hyperbolic media exhibit negative refraction at an interface with elliptic media[29,45,46] (Fig. 4a). Indeed, parallel momentum conservation at the interface (represented by a dotted line in Fig. 4a), combined with the curvature inversion between isotropic and hyperbolic contours, results in a reflected wave and a negatively refracted one (green arrows denote energy flow in Fig. 4a). This configuration can be implemented using our bilayer metasurface with aligned top and bottom lattices ($\theta = 0°$) in the hyperbolic phase, interfaced with an unloaded plate forming the isotropic medium (Fig. 4b). Simulations (left) and experimental results (right) demonstrate (Fig. 4c) negative refraction-mediated focusing of the field emitted by a point source placed in the hyperbolic medium, and its back-reflections.

Negative refraction at an interface has also been recently demonstrated using Weyl metamaterials[47]. Remarkably, due to their topological features it was shown that their dispersion features truncated hyperbolic contours with only half of the branches available, resulting in *reflectionless* negative refraction at the interface with an elliptic medium. Remarkably, our twist-induced hyperbolic shear waves yield an analogous effect: here the strongly asymmetric damping of the IFCs extends the propagation of one hyperbolic branch (green arrows in Fig. 4d), while dramatically hampering the other one (red arrows in Fig. 4d), supporting negative refraction at the interface without reflections. We verify this prediction with the setup in Fig. 4e, where loss asymmetry is induced with a $\theta = 30°$ twist between the two layers (inset). In this scenario, the field emitted by the point source only propagates along one of the hyperbolic branches before negatively refracting at the interface, while the reflected wave is strongly attenuated (Fig. 4f). This additional demonstration further highlights the exciting potential of twist-induced shear phenomena for extreme wave manipulation.

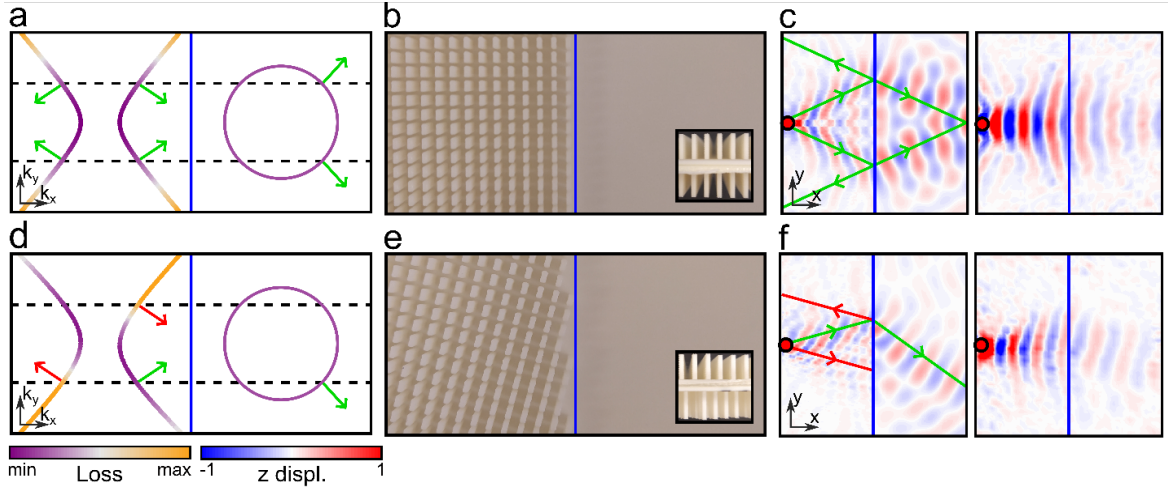

*Figure 4: Shear-induced reflectionless negative refraction. **a,** Momentum matching at the interface between a hyperbolic (left) and an isotropic medium (right). Parallel momentum conservation at the interface results in reflected and negatively refracted waves (green arrows). **b,** Metasurface interface sample corresponding to (**a**): pillars are aligned on both plate sides (inset) on the left of the interface, while the plate is nude on the right. **c,** Pillar metasurface simulation (left) and experimental measurements (right) showing the refocusing of waves emitted by a point source (red disk) across the interface in (**b**), accompanied by back-reflections (energy flux is shown as green arrows). **d,** In the presence of shear, only one set of waves on the right hyperbolic branch reaches the interface (green arrow), and the reflected waves are heavily damped (red arrow on left branch). **e,** Metasurface interface sample corresponding to (**d**) with a twist angle $\theta = 30°$ (inset) for hyperbolic shear waves. **f,** Simulation (left) and measurement (right) of the sample in (**e**), showing reflectionless negative refraction.*

**Axial dispersion-based angular scanning for diffraction-free non-destructive testing**

We conclude by demonstrating the potential of shear elastic metasurfaces for technological applications. In elasticity, a key technological playground is the wave-based detection of mechanical defects, a field broadly referred to as non-destructive testing. Here we show that the combination of extreme directionality and giant axial dispersion exhibited by our hyperbolic shear metasurfaces opens the opportunity for a new paradigm of non-destructive testing. In a conventional mechanical defect detection setup, a material must be scanned by varying the position of the source and detectors used to probe its integrity. However, elliptical wave propagation is notoriously prone to diffraction effects which can spoil the detected signal, and hence fail to accurately detect the location of a defect. Hyperbolic waves can overcome this limitation, thanks to their extreme directionality and ultra-short wavelengths. However, a conventional hyperbolic material only supports hyperbolic propagation along one direction. A shear hyperbolic metasurface, however, can leverage the frequency-dependence of its symmetry axis to steer hyperbolic waves by solely varying the source frequency, which directly encodes their direction of propagation. Furthermore, this enables ultrafast single-shot measurements, whereby an ultrafast pulse can attain all angular information about the location of any defects on a surface, which can be scanned by e.g. coupling it to a shear elastic monoclinic metasurface.

We demonstrate this concept in Fig. 5, where panel a shows a defect-loaded experimental sample. The source is located at the center of the loaded plate, while the defect is clearly visible as the area where the pillars have been removed (see Fig. S5 for details). Panel b shows (top) simulated and (bottom) experimentally measured field maps of the vertical displacement of the plate at different frequencies (left to right). The defect location is marked with a semi-transparent white circle. Note the large steering of the hyperbolic beam, which spans an angle of $\approx 45°$ within a frequency scan of only 8% of the center frequency $f_m$ of the hyperbolic band (see Fig. S6 for details). Panel c shows the (top) theoretical and

(bottom) experimental radiation patterns measured along the red circle in panel a for different frequencies (blue to green, see panel b), clearly showing the shadow created by the defect. Note the contrast between the unobstructed patterns in the lower left quadrant, which evenly scan the entire quadrant, and those in the upper right one, where the defect is located, and the remarkable agreement between simulations and experiment. This is but one of vast range of technological opportunities enabled by hyperbolic shear waves, whose directionality can be encoded in the operating frequency, and hence dynamically steered without any mechanical motion or bias, a concept that can be readily translated to airborne acoustics, as well as photonics and other wave sciences for future sonar and radar applications.

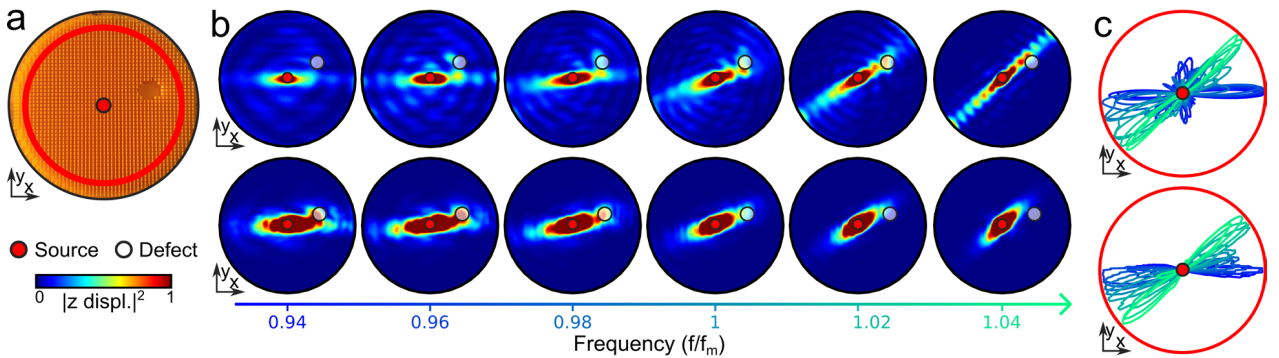

**Figure 5: Diffraction-free non-destructive defect detection based on axial dispersion. a,b,** *We apply the giant axial dispersion of our twisted hyperbolic metasurfaces ($\theta = 30°$) to precisely detect the location of small defects in transmission, with a single, fixed point source, via frequency steering, without suffering from diffraction. In our experimental sample (a), we realize a localized defect by removing a few pillars on both sides. Panel (b) shows (top) simulated and (bottom) experimental field intensity maps for different excitation frequencies, demonstrating the broad steering of the flexural waves by $\approx 45°$ in the canalization regime. This can precisely locate the defect (semi-transparent white circle) without suffering from diffraction, thanks to the ray-like propagation of the hyperbolic*

waves. ***c,*** *Simulated (top) and measured (bottom) radiation patterns across the canalization frequency regime (green to blue), obtained by averaging the field intensities on the red-shaded area in **(a)** for each angle, clearly showing the shadow created by the defect.*

**Conclusions**

In this work we have demonstrated hyperbolic shear waves with giant axial dispersion and maximum shear phenomena, enabled by twisted bilayer metasurfaces. Our model highlights how the interplay of twistronics, non-Hermiticity and extreme anisotropy endows these meta-structures with new forms of wave propagation and loss redistribution. Notably, we demonstrated that maximizing the loss asymmetry by simply tuning the twist angle allows us to drastically extend hyperbolic wave propagation far beyond the distance expected for a given level of material loss. Furthermore, we demonstrated applications of shear metasurfaces to achieve extraordinary wave phenomena such as reflectionless negative refraction, as well as a potential new paradigm for non-destructive testing based on axial dispersion. While we demonstrated these concepts in the elastodynamic domain, our paradigm for shear hyperbolic waves is rooted in the breaking of symmetry through twisting and is therefore extendable to a wide range of metamaterial structures and wave phenomena. We believe that our results open a pathway towards twistronics to harness the combined effect of anisotropy and dissipation for highly directional, long-range wave control in artificial media.

## Methods

### 1. Sample fabrication and experimental setup

Mechanical samples, depicted in Fig. S1(a,b), are 3D printed with polylactic acid (PLA, $E \approx 2.5 GPa$, $\eta \approx 0.3$, $\rho \approx 1300 kg.m^{-3}$) using the Fused Deposition Modeling (FDM) technique (3D printer model is Raise3D Pro2 Plus). The out-of-plane displacement field maps of the bilayer metasurface are measured every 10 Hz between 2 kHz and 12 kHz thanks to a 3D laser vibrometer (Polytec PSV-500-3D), as presented in Fig. S1c. The medium is excited close to its center thanks to a mechanical shaker (B&K type 4810) and a pointer (Fig. S1d). Bidimensional spatial FTs of the field maps allow to extract IFCs and their corresponding loss distribution from the experimental data. The experimental hyperbolic band covers the frequencies between $f_1 \approx 8.8 kHz$ to $f_1 \approx 11 kHz$. Albeit slightly higher in frequency than the pillar system simulations, due to sample fabrication inaccuracies, the results are in good agreement.

2. **Simulation protocol**

The simulation results of pillared metasurface presented in this paper were obtained with the frequency response module of COMSOL Multiphysics Solid Mechanics. Perfectly Matched Layers (PML) are displayed around the 50x50 unit cell sample in order to limit reflections. An out-of-plane displacement source is placed at the center of the system, on the bottom side, in order to excite efficiently the flexural waves in the plate. At each operating frequency, the corresponding out-of-plane displacement field maps on the bottom side of the plate is recorded and a 2D spatial FT is applied to highlight the corresponding IFC. A similar protocol was implemented for the homogenized metasurface cases, using the Mathematics module of COMSOL Multiphysics to solve the corresponding PDE in the case of a point source excitation (Fig. S3).

3. **Theoretical model of general homogenized shear hyperbolic metasurface**

Here we provide a simple derivation of the model demonstrated in Fig. 1 for the case of magnetostatics. We start from Gauss's law for an isotropic medium:

$$\nabla \cdot (\underline{\underline{\varepsilon}} \mathbf{E}) = \nabla \cdot \underline{\underline{\varepsilon}} (-\nabla \varphi - \frac{\partial \mathbf{A}}{\partial t}) = 0, \qquad (3)$$

Where we substituted the standard expression for the electric field in terms of the scalar and vector electromagnetic potentials $\varphi$ and $\mathbf{A}$. We now apply the anisotropic form of the Lorenz gauge[48]:

$$\nabla \cdot (\underline{\underline{\varepsilon}} \mathbf{A}) = -\frac{\partial \phi}{\partial t} \tag{4}$$

and substitute the left-hand side into Eq.(3). Assuming, that $\varepsilon$ is constant in time, it commutes with the time derivative, so that Eq.(4) becomes:

$$\nabla \cdot \underline{\underline{\varepsilon}} \nabla \varphi - \frac{\partial^2 \varphi}{\partial t^2} = 0, \tag{5}$$

Interestingly, this model coincides exactly with the Kirchoff-Love case, if the nonlocal bilaplacian term is ignored. To evaluate loss, in our elasticity calculations we compute the scalar product $\nabla w^\dagger \cdot \mathfrak{I}[\hat{\tau}^\dagger] \cdot \nabla w$ (where $w$ is the flexural displacement of the plate), which quantifies the effect of resonant loss within the medium, paralleling the standard expression $\mathbf{E}^\dagger \mathfrak{I}[\hat{\varepsilon}^\dagger] \mathbf{E}$ used for power loss rate in electrodynamics.

### 4. Derivation of axial dispersion and shear factor

Diagonalization of $\mathfrak{R}[\hat{\tau}]$ gives the angle $\beta$ that its two eigenvectors form with the $x$ and $y$ axes:

$$\beta = \frac{1}{2} \arctan\left( \frac{\mathfrak{R}[\tau_2]\sin(2\theta)}{\mathfrak{R}[\tau_1] + \mathfrak{R}[\tau_2]\cos(2\theta)} \right) \tag{6}$$

Rotating the entire $\hat{\tau}$ tensor by $\beta$ gives:

$$\hat{\tau}' = \begin{pmatrix} \tau'_{xx} & \tau'_{xy} \\ \tau'_{yx} & \tau'_{yy} \end{pmatrix} = \begin{pmatrix} \mathfrak{R}[\tau'_{xx}] & 0 \\ 0 & \mathfrak{R}[\tau'_{yy}] \end{pmatrix} + i \begin{pmatrix} \mathfrak{I}[\tau'_{xx}] & \mathfrak{I}[\tau'_{xy}] \\ \mathfrak{I}[\tau'_{yx}] & \mathfrak{I}[\tau'_{yy}] \end{pmatrix} \tag{7}$$

With:

$$\begin{aligned} \tau'_{xx} &= \tau_1 \cos^2(\beta) + \tau_2 \cos^2(\beta - \theta) \\ \tau'_{yy} &= \tau_1 \sin^2(\beta) + \tau_2 \sin^2(\beta - \theta) \\ \tau_{xy}' &= \{\sin(2\beta)\mathfrak{I}[\tau_1] + \sin[2(\beta - \theta)]\mathfrak{I}[\tau_2]\}/2 \end{aligned} \tag{8}$$

$\tau_{xy}'$ can be normalized to 1 to obtain the shear coefficient:

$$S(\omega,\theta) = \frac{\tau_{xy'}}{N^I} = \frac{\sin(2\beta)\Im[\tau_1] + \sin[2(\beta-\theta)]\Im[\tau_2]}{\Im[\tau_1] + \Im[\tau_2]} \quad (9)$$

plotted in Fig. 3b of the main text, where we defined the normalization factor $N^I = (\Im[\tau_1] + \Im[\tau_2])/2$.

**Methods references**

**Acknowledgements**

S.Y., X.N., E.M.R. and A.A. acknowledge funding from the Simons Foundation. E.G acknowledges funding from the Simons Foundation through a Junior Fellowship of the Simons Society of Fellows (855344, EG).

**Data availability**

The data that support the findings of this study are available from the corresponding authors upon reasonable request.

**Code availability**

All codes used to produce and analyze data that support the findings of this study are available from the corresponding authors upon reasonable request.

**Competing interests**

The authors declare no competing interests.

**Contributions**

SY carried out the experiments, and EG contributed to data analysis. EG led the analytic theory, with contributions from SY, XN, and EMR. SY, XN carried out finite element simulations, with support from EMR. SY and EG wrote the manuscript, with contributions from all co-authors. AA devised and supervised the project. All co-authors discussed the development and results of the project.

**Ethics declarations**

N/A

**Supplementary information**

Supplementary Information is available for this paper.